\documentclass[lettersize,journal]{IEEEtran}
\usepackage{amsmath,amsfonts}
\usepackage{array}
\usepackage{bm}
\usepackage{textcomp}
\usepackage{stfloats}
\usepackage{url}
\usepackage{verbatim}
\usepackage{graphicx}
\usepackage{cite}
\usepackage{booktabs}
\usepackage{makecell}
\usepackage[flushleft]{threeparttable}
\newcolumntype{C}[1]{>{\centering\arraybackslash}p{#1}}
\usepackage[utf8]{inputenc} %
\usepackage[T1]{fontenc}    %
\usepackage[colorlinks,
    linkcolor=red,citecolor=citecolor,urlcolor=blue]{hyperref}       %
\usepackage[table,xcdraw]{xcolor}

\usepackage{algorithm}
\usepackage{graphicx}
\usepackage{threeparttable}
\usepackage{multirow}
\usepackage{amsmath}
\usepackage{bm}
\usepackage{bbm}
\usepackage{amsthm}
\usepackage[]{algpseudocode}
\usepackage{enumitem} %
\usepackage[subrefformat=parens,labelformat=parens]{subfig}

\usepackage{wrapfig}
\usepackage[algo2e, ruled, vlined, linesnumbered, noend]{algorithm2e}

\graphicspath{{./figs/}}
\newcommand{\doctor}{\texttt{DOCTOR}\xspace}
\definecolor{citecolor}{RGB}{34,139,34}
\definecolor{mydarkblue}{rgb}{0,0.08,1}
\definecolor{mydarkgreen}{rgb}{0.02,0.6,0.02}
\definecolor{mydarkred}{rgb}{0.8,0.02,0.02}
\definecolor{mydarkorange}{rgb}{0.40,0.2,0.02}
\definecolor{mypurple}{RGB}{111,0,255}
\definecolor{myred}{rgb}{1.0,0.0,0.0}
\definecolor{mygold}{rgb}{0.75,0.6,0.12}
\definecolor{myblue}{rgb}{0,0.2,0.8}
\definecolor{mydarkgray}{rgb}{0.,0.2,0.2}

\definecolor{lightred}{RGB}{255,235,235}
\definecolor{lightgreen}{RGB}{235,255,235}
\definecolor{lightblue}{RGB}{235,235,255}
\definecolor{lightcyan}{RGB}{235,255,255}
\definecolor{lightmagenta}{RGB}{255,235,255}
\definecolor{lightyellow}{RGB}{255,255,235}

\definecolor{qxkcolor}{RGB}{215,235,255}
\definecolor{softmaxcolor}{RGB}{230,235,255}
\definecolor{probxvcolor}{RGB}{255,255,235}

\definecolor{topkcolor}{RGB}{255,235,235}
\definecolor{zecolor}{RGB}{255,255,235}
\definecolor{dynacolor}{RGB}{235,255,255}

\definecolor{reviewcolor}{RGB}{0,0,200}

\newcommand{\ceil}[1]{\lceil #1 \rceil}

\newcommand{\calL}{\mathcal{L}}

\newcommand{\E}{\mathbb{E}}

\theoremstyle{plain}

\theoremstyle{definition}

\algdef{SE}[DOWHILE]{Do}{doWhile}{\algorithmicdo}[1]{\algorithmicwhile\ #1}%

\newcommand{\squishlist}{
 \begin{list}{$\bullet$}
  { \setlength{\itemsep}{0pt}
     \setlength{\parsep}{3pt}
     \setlength{\topsep}{3pt}
     \setlength{\partopsep}{0pt}
     \setlength{\leftmargin}{1.5em}
     \setlength{\labelwidth}{1em}
     \setlength{\labelsep}{0.5em} } }
     
\newcommand{\squishend}{
  \end{list}  }

\begin{document}
\title{
DOCTOR: \underline{D}ynamic \underline{O}n-\underline{C}hip Remediation Against \underline{T}emporally-Drifting Thermal Variations Toward Self-Corrected Ph\underline{o}tonic Tensor Accele\underline{r}ators
}

\title{
DOCTOR: Dynamic On-Chip Temporal Variation Remediation Toward Self-Corrected Photonic Tensor Accelerators
}

\author
{
Haotian Lu,
Sanmitra Banerjee,
Jiaqi Gu,~\IEEEmembership{IEEE Member}\\
Arizona State University\\
\small\textit{jiaqigu@asu.edu}
}

\markboth{JOURNAL OF LIGHTWAVE TECHNOLOGY}%
{Shell \MakeLowercase{\textit{et al.}}: A Sample Article Using IEEEtran.cls for IEEE Journals}

\maketitle

\begin{abstract}
\label{abstract}
Photonic computing has emerged as a promising solution for accelerating computation-intensive artificial intelligence (AI) workloads, offering unparalleled speed and energy efficiency, especially in resource-limited, latency-sensitive edge computing environments. 
However, the deployment of analog photonic tensor accelerators encounters reliability challenges due to hardware noise and environmental variations.
While off-chip noise-aware training and on-chip training have been proposed to enhance the variation tolerance of optical neural accelerators with \emph{moderate}, \emph{static} noise, we observe a notable performance degradation over time due to \emph{temporally drifting} variations, which requires a real-time, in-situ calibration mechanism. 
To tackle this challenging reliability issues, \emph{for the first time}, we propose a lightweight dynamic on-chip remediation framework, dubbed \doctor, providing adaptive, in-situ accuracy recovery against temporally drifting noise.
The \doctor framework intelligently monitors the chip status using adaptive probing and performs fast in-situ training-free calibration to restore accuracy when necessary. 
Recognizing nonuniform spatial variation distributions across devices and tensor cores, we also propose a variation-aware architectural remapping strategy to avoid executing critical tasks on noisy devices.
Extensive experiments show that our proposed framework can guarantee sustained performance under drifting variations with 34\% higher accuracy and 2-3 orders-of-magnitude lower overhead compared to state-of-the-art on-chip training methods.
Our code is open-sourced at \href{https://github.com/ScopeX-ASU/DOCTOR}{link}.
\end{abstract}


\section{Introduction}
\label{sec:Introduction}

In recent years, the pursuit of efficient and high-performance solutions for artificial intelligence (AI) workloads has led to the emergence of photonic computing. 
Leveraging the unique properties of light, analog photonic accelerators stand out for their ability to deliver unparalleled speed and efficiency, presenting a promising avenue for AI applications~\cite{NP_NATURE2017_Shen, NP_PIEEE2020_Cheng, NP_NaturePhotonics2021_Shastri, NP_ACS2022_Feng, NP_DATE2019_Liu, NP_APLML2024_Gu, NP_Nature2021_Xu, NP_Nature2021_Feldmann, NP_APLPhotonics2020_Huang}.

\begin{figure}[]
    \centering
    \subfloat[]{
    \includegraphics[width=0.365\columnwidth]{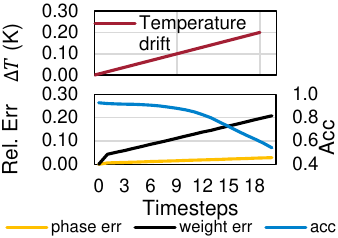}
    \label{fig:Motivation_Drift}
    }
    \subfloat[]{
    \includegraphics[width=0.3\columnwidth]{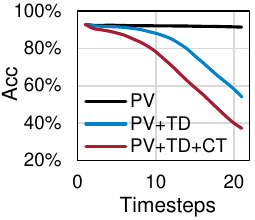}
    \label{fig:Motivation_Variation}
    }
    \subfloat[]{
    \includegraphics[width=0.203\columnwidth]{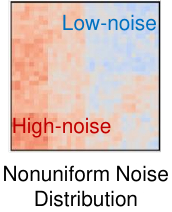}
    \label{fig:Motivation_Distribution}
    }
    \caption{\small (a) MRR-based photonic accelerator is sensitive to temperature drift.
    (b) Drifting noises cause severe accuracy drop over time, including phase variation (PV), temperature drift (TD), and thermal crosstalk (CT). "Acc" represents the accuracy.
    (c) Noise distributions across devices are nonuniform.}
    \label{fig:Motivation}
\end{figure}

However, the deployment of such accelerators encounters robustness challenges that impede their practical application~\cite{NP_DATE2020_Gu, NP_ICCAD2019_Zhao, NP_ICCAD2020_Zhu,NP_TCAD2022_Mirza}.
We consider one of the most sensitive accelerators based on micro-ring resonators (MRRs) as a case study~\cite{NP_SciRep2017_Tait, NP_DAC2018_Liu, NP_APLPhotonics2020_Huang}. 
Due to the intrinsic temperature sensitivity of the MRR device, shown in Fig.~\ref{fig:Motivation_Drift}, a subtle drift in the temperature will lead to a slight change of the round-trip phase shift but a large deviation on the represented weight.
Such a high sensitivity makes pre-deployment optimization ineffective and thus necessitates a real-time calibration mechanism on chip. 
Besides temperature drift, various dynamic random noise and crosstalk cast even more shadows on the reliability of photonic computing systems.
Figure~\ref{fig:Motivation_Variation} shows the significant impacts of variations on accuracy, sometimes leading to malfunction over time when the noise intensities gradually increase.

While previous off-chip noise-aware model training~\cite{NP_DATE2020_Gu, NP_DATE2021_Gu2} have shown efficacy in enhancing the variation tolerance of optical accelerators by injecting noise during training and thus encourage a smoother solution space, they rely on accurate noise modeling and thus typically show unsatisfying robustness improvement with unknown physical variations and can only handle \emph{small and static} noise~\cite{NP_DATE2020_Gu, NP_ICCAD2019_Zhao,NP_TETCI2023_Kirtas,NP_Nature2022_Wright}.
The performance drop remains unresolved when there exist temporally drifting variations.
Recently, there has been a trend to resort to on-chip learning or physical training methods to directly train the optical neural network (ONN) models \emph{in situ} that can naturally incorporate real physical noise into the weight training process~\cite{NP_DAC2020_Gu, NP_AAAI2021_Gu,NP_NeurIPS2021_Gu, NP_Nature2022_Wright,NP_Science2023_Pai}.
However, they require repeatedly performing forward and backward propagation of the entire network on a labeled training dataset to calculate the task-specific gradients for weight fine-tuning, which induces nontrivial training costs that can severely harm the system throughput and efficiency.
Moreover, prior methods fail to leverage the nonuniform spatial noise distribution and weight sensitivity, shown in Fig.~\ref{fig:Motivation_Distribution}, to balance accuracy and efficiency.
Hence, it necessitates a \emph{real-time, low-cost, in-situ} calibration mechanism without running backpropagation on any labeled training set to quickly recover the accuracy and ensure \emph{continued reliability} in practical deployment.

To tackle these challenges, we present a \doctor framework for dynamic on-chip remediation against temporally drifting thermal variations. 
In this paper, we delve into the detailed modeling of time-variant thermal variations and their impacts on a thermal-sensitive MRR-based photonic accelerator and resolve the variation-induced performance drop by efficient sparse weight calibration, variation-aware tile remapping, and an adaptive remediation controller.

The major contributions of this paper are as follows:

\squishlist
    {\item \textbf{Thermal Variation Modeling}: We give rigorous modeling and sensitivity analysis of the dynamic thermal variations for multi-core photonic accelerators, providing a deeper understanding of the dynamic variations in real-world deployment.}
    {\item \textbf{Salience-Aware Sparse Calibration}: We propose a training-free, data-free \emph{in-situ} calibration mechanism to selectively 
    mitigate thermal variations and effectively resume computing accuracy at negligible runtime overhead.
    }
    {\item \textbf{Variation-Aware Tile Remapping}: We leverage the spatial nonuniformity in noise distributions to boost the reliability by optimally remapping workloads onto tensor cores, aware of the weight importance and device noise levels.}
    {\item We evaluate that our \doctor framework guarantees sustained deployment performance with 1\%-2.5\% accuracy drop at negligible runtime overhead (0.1\%-5\% cycle overhead), outperforming state-of-the-art on-chip training methods~\cite{NP_NeurIPS2021_Gu, NP_Science2023_Pai} by an average of 34\% higher accuracy and 2-3 orders-of-magnitude less overhead. 
    Our work makes significant strides toward the real-world deployment of photonic accelerators in dynamic environments.}
\squishend

\section{Background}
\label{sec:Background}
\subsection{Photonic Tensor Accelerators}
\label{sec:PhotonicsBasics}

Various photonic neural network designs have been proposed and demonstrated to encode inputs and weights to the light magnitude/phase and circuit transmission, respectively, and perform ultra-fast matrix multiplication~\cite{NP_NATURE2017_Shen, NP_PIEEE2020_Cheng, NP_NaturePhotonics2021_Shastri, NP_ACS2022_Feng, NP_DATE2019_Liu}.
Typically, the photonic circuits are sensitive to thermal variation as temperature impacts the refractive index of the optical component.
Especially for compact microring resonator (MRR)-based photonic accelerators, the weight $w$ is encoded by the differential transmission of the add-drop MRR as $w=g(2a-1);~a=\frac{\alpha^2-2r\alpha \cos{\phi}+r^2}{1-2r\alpha \cos{\phi}+r^2\alpha^2}\in(0,1)$,
where $\alpha$ is the attenuation factor, $r$ is the coupling factor, $\phi$ is the round-trip phase shift, $a$ is the through-port transmission, and $g$ is the scaling factor.
To accommodate multiple channels in a single free-spectral range (FSR) while minimizing spectral crosstalk, i.e., large finesse, MRR is designed to have a mid-to-high quality factor.
This design choice makes the MRR highly sensitive to small disturbances, highlighting the urgency in addressing challenges related to thermal variation robustness.

\begin{figure}
    \centering
    \includegraphics[width=\columnwidth]{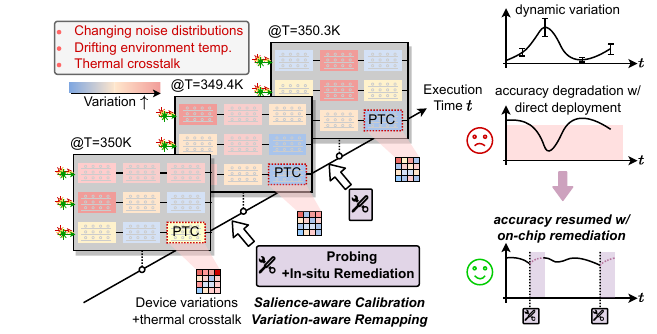}
    \caption{\small Our proposed dynamic remediation \doctor can counter the accuracy degradation due to temporally drifting hardware variations.}
    \label{fig:Overview}
\end{figure}

\begin{figure*}[]
    \centering
    \includegraphics[width=0.95\textwidth]{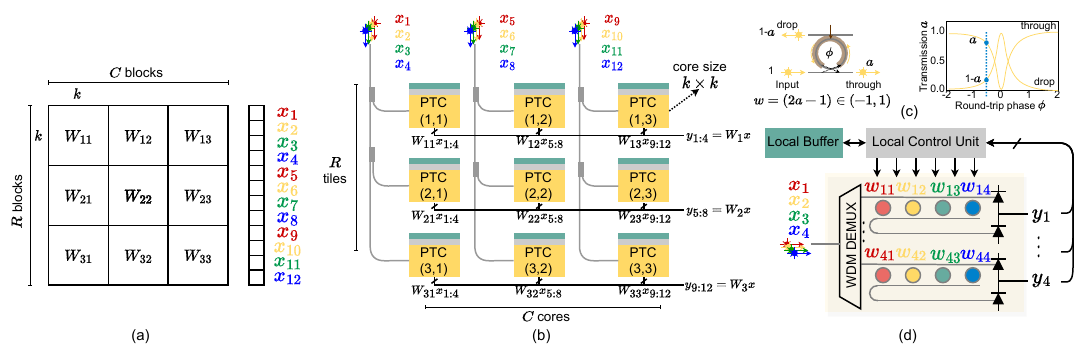}
    \caption{\small Architecture settings of a multi-core photonic tensor accelerator.
    (a) The accelerator can map $Rk\times Rk$ matrix-vector multiplication at each cycle.
    Note that we draw $R=C=3$ and $k=4$ as an example for illustration but not the actual architecture setting.
    (b) The photonic accelerator include $R$ tiles, each tile including $C$ photonic tensor cores (PTCs). Each PTC is of size $k \times k$.
    Partial sum accumulation is performed by photocurrent accumulation across $C$ cores within one tile.
    The same input vector chunks are broadcast to $R$ tiles (vertically) using photonic on-chip interconnects.
    (c) An ideal add-drop micro-ring resonator has a tunable through-port transmission $a$ and a corresponding drop-port transmission $1-a$.
    (d) As a case study, each $k\times k$ PTC is assumed to be a multiple-wavelength add-drop MRR weight bank with local buffers and electronic local control units.}
    \label{fig:ArchSetting}
    \vspace{-5pt}
\end{figure*}

\subsection{Noise-Aware ONN Optimization}
\label{sec:RobustnessONN}
Noise-aware optimization for photonic accelerators includes two categories, i.e., offline optimization and on-chip optimization.
Prior offline methods inject noise into ONN model training to obtain smooth solution space for better noise tolerance~\cite{NP_DATE2020_Gu}.
However, such a method requires accurate noise modeling and can only handle static, known variations, which cannot be adapted to dynamically drifting, unknown variations.
On-chip training, as a trend, has been demonstrated for online adaptability and efficiency in \emph{in-situ} accuracy recovery.
The pretrained optical NNs are fine-tuned on a target dataset with on-chip gradient calculation~\cite{ NP_DAC2020_Gu, NP_AAAI2021_Gu, NP_NeurIPS2021_Gu}.
However, they rely on accurate gradient calculation and require costly forward and backward propagation on a given training set, which can significantly degrade the edge inference throughput and efficiency and bear data privacy issues.

\section{Dynamic On-Chip Remediation: \doctor}
\label{sec:Method}
We introduce our efficient on-chip remediation flow \doctor, shown in Fig.~\ref{fig:Overview}, with rigorous noise modeling, lightweight device calibration, and architectural remapping techniques to guarantee real-time accuracy recovery against drifting variations.

\subsection{Photonic Accelerator Architecture Settings}
\label{sec:ArchSetting}
Since MRR weight banks are typically considered to be one of the most thermal sensitive designs among different kinds of ONN designs~\cite{NP_TVLSI2013_Ye,NP_Nanophotonics2014_Padmaraju, NP_JLT2019_Milanizadeh}, we focus on photonic accelerator architectures based on MRR weight banks as a challenging case study to showcase the effectiveness of our \doctor method, shown in Fig.~\ref{fig:ArchSetting}.
We assume a multi-core accelerator with $R$ tiles and $C$ photonic tensor cores (PTCs) per tile.
Each PTC is a $k\times k$ add-drop MRR weight bank.
The partial sums are reduced in each tile.
Hence, it can finish an $Rk\times Ck$ matrix-vector multiplication (MVM) at each cycle.
A large $M\times N$ matrix will be partitioned and simply mapped to this accelerator using $\ceil{M/(Rk)}\cdot \ceil{N/(Ck)}$ cycles.
Alongside each PTC, we assume a dedicated local buffer, an electrical local control unit, and thermal sensors for temperature monitoring, control, and processing.

\subsection{Thermal Variation Modeling}
\label{sec:ThermalVariationModeling}
We first give a thorough noise modeling and sensitivity analysis for this photonic accelerator, including dynamic phase variations, drifting environmental temperature, and inter-device thermal crosstalk.
\subsubsection{Device Phase Variation}
\begin{figure}
    \centering
    \includegraphics[width=0.99\columnwidth]{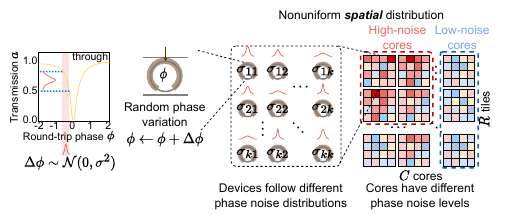}
    \caption{\small Random phase variations on MRRs lead to large weight errors.
    Different devices and cores have distinct noise distributions.}
    \label{fig:PhaseNoise}
    \vspace{-10pt}
\end{figure}

Due to control signal noise and thermal fluctuations, the round-trip phase shift of MRRs exhibits stochastic variations. 
In Fig.~\ref{fig:PhaseNoise}, we posit a zero-mean Gaussian phase variation, denoted as $\Delta\phi\sim\mathcal{N}(0,\sigma^2)$~\cite{NP_OE2019_Fang}. 
This introduces a noisy transmission and weight values. 
Given that different MRR devices are subject to distinct noise distributions due to their unique control sources and local environments, our figure depicts an instance where the upper-left corner of the accelerator experiences more noise, in contrast to the lower-right corner of the chip, which exhibits less noise.

\begin{figure}
    \centering
    \includegraphics[width=0.99\columnwidth]{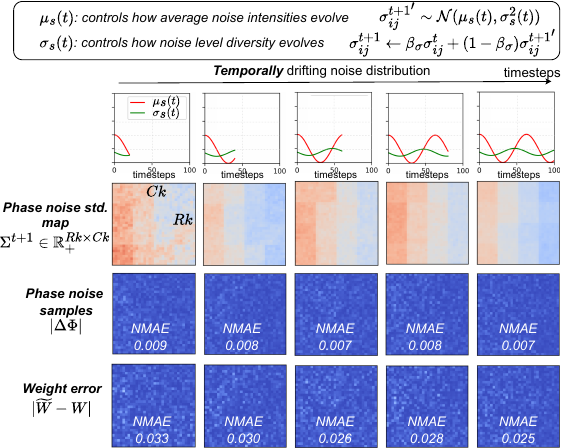}
    \caption{\small Illustration of temporally drifting phase noise distributions.
    We control the mean and std of the distribution.
    At every timestep, it samples a new noise std map from the scheduled distribution and smoothly evolves to a new distribution via a damping factor $\beta_{\sigma}$.
    }
    \label{fig:PhaseNoiseDrift}
\end{figure}

Beside nonuniform noise distribution, the noise profiles are dynamically drifting over time, as shown in Fig.~\ref{fig:PhaseNoiseDrift}.
We simulate such dynamics with two-level sampling.
For each PTC, we use the Standard Deviation ($\sigma_{ij}$, std) of the phase noise distribution to represent the noise intensity for the MRR at coordinate $ij$, while the intensity itself is time-varying.
At time step $t$, we sample the step-$t$ noise intensity from a distribution, i.e., $\sigma_{ij}^{t+1'} \sim \mathcal{N}(\mu_s(t),\sigma_s^2(t))$.
We apply exponential moving average $\sigma_{ij}^{t+1}\gets \beta_{\sigma}\sigma_{ij}^t+(1-\beta_{\sigma})\sigma_{ij}^{t+1'}$ with a dampling factor $\beta_{\sigma}$ to smooth out the intensity drifting.
$\mu_s(t)$ and $\sigma_s(t)$ are temporally evolving with a certain scheduling function.

The non-uniform, temporally drifting spatial distribution of noise provides valuable insights into addressing this challenge through periodic workload-to-PTC remapping, which implies intuitively mapping sensitive workloads to less noisy tensor cores. 

\subsubsection{Environmental Temperature Drift}
\begin{figure}
    \centering
    \includegraphics[width=0.99\columnwidth]{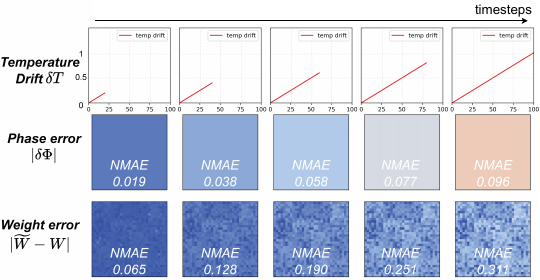}
    \caption{\small Illutration of linear uniform temperature drift with a maximum 1K change, which introduces uniform phase errors and nonuniform weight errors across $Rk\times Ck$ MRRs.}
    \label{fig:TempDrift}
\end{figure}

The environmental temperature fluctuation can impact the reliability of the chip, especially for thermal-sensitive microring resonator-based PTCs~\cite{NP_DATE21_Sunny,NP_TVLSI2013_Ye,NP_Nanophotonics2014_Padmaraju}.
As shown in Fig.~\ref{fig:Motivation_Variation}, for MRRs, even 0.05K temperature drift can lead to a relatively large resonant wavelength shift, leading to large errors on the represented weight value.
We assume a linear dependence of on-resonance wavelength $\lambda_c$ on temperature $T$. Thus, we assume a constant $\partial\lambda/\partial T$.
The drift on the round-trip phase shift $\delta\phi$ due to temperature change is derived as,
\begin{equation}
\small
    \begin{aligned}
    \label{eq:TemperatureDrift}
    \delta\lambda&=\delta T (\partial\lambda / \partial T),~~\delta n_{eff}= \delta\lambda \cdot n_g/\lambda,~~
    \delta\phi=\delta n_{eff}\cdot 2\pi\cdot L/\lambda,
    \end{aligned}
\end{equation}
where $n_g$ is the group index, $\lambda$ is the input wavelength (i.e., ideal on-resonance wavelength at temperature $T_0$), $L$ is the perimeter of the MRR, $n_{eff}$ is the effective refractive index.
For the MRR on the $c$-th column of the weight bank, its input wavelength is $\lambda_c$, and its corresponding perimeter is $L_c$.
$L_c$ is designed to make the MRR resonate at $\lambda_c$ and default temperature $T_0$.
Hence its round-trip phase shift change is $\delta\phi_c=\delta n_{eff}\cdot 2\pi\cdot L_c/\lambda_c$.
Figure~\ref{fig:TempDrift} illustrates a linear temperature drift scheduling from 300K to 301K and its impacts on phase error and weight error measured as normalized mean-absolute error (NMAE).

\subsubsection{Thermal Crosstalk}

\begin{figure}
    \centering
    \includegraphics[width=0.99\columnwidth]{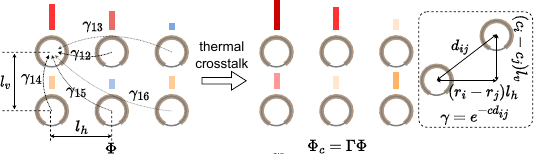}
    \caption{\small Thermal crosstalk among MRRs within the same weight bank.
    The MRR spacings are $l_h$ and $l_v$.
    The coupling matrix $\Gamma$ is applied to all phase shifts $\Phi$.
    The crosstalk factor $\gamma$ exponentially decays with a larger device spacing $d$.}
    \label{fig:ThermalCrosstalk}
\end{figure}

To make a compact MRR-based PTC, the spacings between adjacent rings are usually not far enough to eliminate all thermal crosstalk, which is mainly due to the thermal interference between adjacent photonic devices~\cite{NP_JLT2015_Jayatilleka,NP_LPR2012_Bogaerts,NP_JLT2019_Milanizadeh}.
Figure~\ref{fig:ThermalCrosstalk} illustrates the crosstalk within a weight bank.
Given a fixed layout spacing of MRRs, $\gamma_{ij}$ represents the crosstalk coupling coefficient from the $j$-th MRR to $i$-th MRR,
The round-trip phases $\Phi$ for $k\times k$ MRRs will be transformed with a coupling matrix $\Gamma$~\cite{NP_JLT2019_Milanizadeh}, i.e., $\Phi_c=\Gamma\Phi$, as follows,
\begin{equation}
\small
    \label{eq:CrosstalkMatrix}
    \begin{aligned}
        \begin{pmatrix}
        \phi^c_1\\
        \phi^c_2\\
        \vdots\\
        \phi^c_{k^2}
    \end{pmatrix}&=\begin{pmatrix}
        \gamma_{11} & \gamma_{12} & \cdots & \gamma_{1k^2}\\
        \gamma_{21} & \gamma_{22} & \cdots & \gamma_{2k^2}\\
        \vdots      & \vdots      & \ddots & \vdots\\
        \gamma_{k^21} & \gamma_{k^22} & \cdots & \gamma_{k^2k^2}
    \end{pmatrix}
    \begin{pmatrix}
        \phi_1\\
        \phi_2\\
        \vdots\\
        \phi_{k^2}
    \end{pmatrix},\\~~
    \gamma_{ii}&=1,~~\gamma_{ij}=e^{-k_1 d_{ij}},\\
    d_{ij}&=\sqrt{\big((r_j-r_i)l_v\big)^2+\big((c_j-c_i)l_h\big)^2},~~r_{i/j}\in[k], c_{i/j}\in[k],
    \end{aligned}
\end{equation}
where $\gamma_{ii}=1$ is the self-coupling coefficient, and $\gamma_{ij}=e^{-k_1 d_{ij}}\in(0,1)$ is the cross-coupling coefficient, determined by the structure-related constant $k_1$ and the distance $d_{ij}$ between the $i$-th MRR at coordinate ($r_i,c_i$) and $j$-th MRR at coordinate ($r_j,c_j$)\cite{NP_IPC2023_Cem}.

\subsection{Salience-Aware Sparse Calibration}

\begin{figure}
    \centering
    \includegraphics[width=\columnwidth]{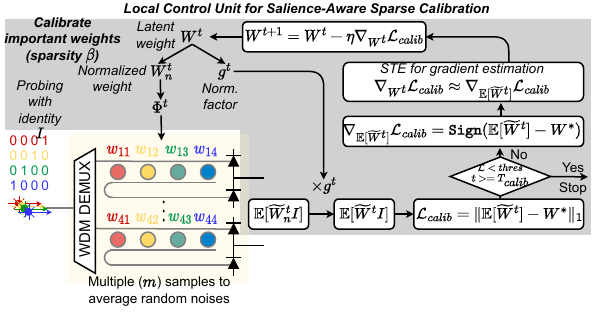}
    \caption{\small The proposed sparse calibration flow performs backpropagation-free data-free local regression.}
    \label{fig:Calibration}
\end{figure}

Given a pretrained NN, we map the ideal weights $W^*$ to the photonic accelerator.
However, the actual weights $\widetilde{W}$ will have a deviation from the ideal ones.
To quickly recover the inference accuracy on the fly without training or utilizing any labeled calibration datasets, we introduce salience-aware sparse calibration in Fig.~\ref{fig:Calibration}, which is formulated as a batched block-wise regression problem,
\begin{equation}
\small
\begin{aligned}
    \label{eq:Calibration}
    \min_{W} \calL_{calib}&=\min_{W} \sum_{ij}\big|\calL(\E[\widetilde{W}_{ij}])-\calL(W^*_{ij})\big|,\\
    ~~\widetilde{W}_{ij}&=W_{ij}\big(\Gamma(\Phi+\Delta\Phi+\delta\Phi)\big)
\end{aligned}
\end{equation}
By optimizing the latent weight $W$, we want to minimize the distance between the expected noisy weights $\E[\widetilde{W}]$ and the ideal weights $W^*$ on each $k\times k$ block.
The loss degradation of the NN can be approximated by Taylor expansion
\begin{equation}
\small
    \label{eq:LossApproximation}
    \begin{aligned}
         |&\calL\!(\E[\widetilde{W}])\!-\!\!\calL\!(W^*)|\!\approx\! \big|\nabla_{W}\!\calL^T\!(\E[\widetilde{W}]\!\!\!-\!\!W^*) \!+\! \frac{1}{2}\!\nabla^2_{W}\!\calL(\E[\!\widetilde{W}\!]\!\!-\!\!W)^2\big|.
    \end{aligned}
\end{equation}
As an approximation,
we can rewrite the objective $\calL_{calib}$ in Eq.~\eqref{eq:Calibration} 
\begin{equation}
\small
    \label{eq:FirstOrderCalibration}
    \min_{W} \calL_{calib}\approx \min_{W} \sum_{ij}\|\E[\widetilde{W}_{ij}]-W^*_{ij}\|_1.
\end{equation}
This fundamentally \emph{decouples the calibration procedure from the labeled dataset and task-specific loss} function $\calL$.
By solving this regression problem concurrently on all matrix blocks ($\forall i,j$), we can efficiently resume the task performance.
The expected noisy weights $\E[\widetilde{W}]$ can be probed by shining an identity matrix $I\in\mathbb{R}^{k\times k}$ through the MRR weight bank $m$ times and calculating the average,
\begin{equation}
\small
    \label{eq:AverageProbe}
    \E[\widetilde{W}]\approx \frac{1}{m}\sum_{i=1}^m\widetilde{W}I=\frac{1}{m}\sum_{i=1}^m W\big(\Gamma(\Phi+\Delta\Phi_i+\delta\Phi)\big)I.
\end{equation} 
Compared to on-chip training with backpropagation~\cite{ NP_Science2023_Pai, NP_NeurIPS2021_Gu}, our calibration method has three major advantages:

(1) \textbf{High efficiency} -- Our method does not require costly forward propagation or error feedback.
The gradient of the calibration objective w.r.t. the latent weights $\nabla_{W}\calL_{calib}$ can be efficiently approximated using a straight-through estimator.
With MAE as the loss, the gradients are simply
\begin{equation}
\small
    \label{eq:STE}
    \nabla_{W}\calL_{calib}\approx\nabla_{\E[\widetilde{W}]}\calL_{calib}=\texttt{Sign}(\frac{1}{m}\sum_{i=1}^m\widetilde{W}_iI-W^*).
\end{equation}

(2) \textbf{Accurate gradients} -- 
Unlike on-chip training, where gradient errors will exponentially accumulate through layers, our method only calculates gradients for each block without backpropagation.
Hence, our method avoids slow convergence or divergence issues even under large noise. 

(3) \textbf{Task-agnostic \& Data-free} -- We do not use any labeled dataset or task-specific loss function, which avoids dataset storage costs and data privacy issues on edge devices.

\noindent\textbf{Salience-Aware Sparsity}.~
To reduce the overhead of calibration, we propose salience-aware sparse calibration.
At each iteration, we only calibrate a subset ($\beta\times 100$\%) of weight blocks, e.g., $\beta=0.2$ means 20\% sparsity.
Instead of randomly selecting blocks to be calibrated, we propose to prioritize important weights based on salience scores that can be precomputed offline once.
The precomputed first-order or second-order gradients are good indicators of weight importance/salience, i.e., how sensitive the task loss function is to the weight perturbation.
Therefore, we will generate salience scores, e.g., $s=|\nabla_W\mathcal{L}(\mathcal{D}_{train})|$ or $s=|\nabla^2_W\mathcal{L}(\mathcal{D}_{train})|,$ of ideal weights across the training dataset, calculate the average scores for each $Rk\times Ck$ weight chunk, and use the normalized salience score as sampling probability to perform importance-sampling (IS) at each calibration iteration.
This coarse-grained structured sparsity at the chunk level can directly translate to proportional calibration cycle reduction.
We set a maximum calibration iteration and a weight error threshold to adaptively stop calibration, whichever stop criterion is first met.
The hardware overhead in terms of cycles when performing $T_{calib}$ iterations for each $Rk\times Ck$ weight block is,
\begin{equation}
    \label{eq:CalibrationCost}
    \#\texttt{Cycle}_{calib}\approx \beta T_{calib}mk
\end{equation}

\subsection{Variation-Aware Tile Remapping}
\begin{figure*}
    \centering
    \includegraphics[width=0.75\textwidth]{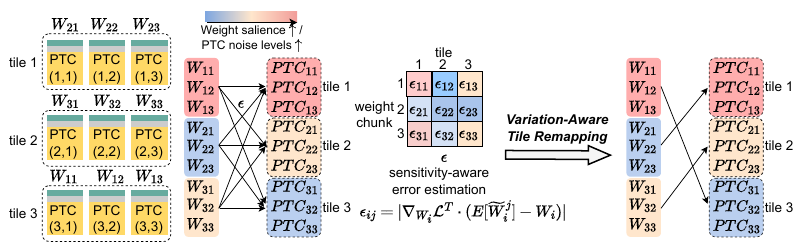}
    \vspace{-5pt}
    \caption{\small The proposed variation-aware tile remapping method first probes the errors for all matrix-to-tile pairs and then solves a linear assignment problem to find the min-error reordering.}
    \label{fig:TileRemapping}
\end{figure*}

Motivated by the important observation of the nonuniform noise distribution across devices and cores in Fig.~\ref{fig:Motivation_Distribution}, we propose architectural variation-aware tile remapping to remedy the accuracy degradation.
We try to find a matrix-to-tile index mapping better than the direct mapping ordering, shown in Fig.~\ref{fig:TileRemapping}.
Given a weight-stationary dataflow, we map a $Rk\times Ck$ weight matrix block onto the accelerator for MVMs and move to the next matrix block.
A direct mapping will map $R \times C$ subblocks onto the $R\times C$ PTC arrays following their original order.
This is suboptimal because weight blocks have different sensitivities, and PTCs show different error levels.
It is natural to remap weight blocks onto PTCs to minimize errors.

However, to avoid complicated dataflow, we cannot arbitrarily remap $RC$ weight blocks to $RC$ PTCs, which will lead to nontrivial architectural overhead.
First, as imposed by the dataflow in the accelerator topology, the same input will be broadcast via photonic waveguides to all cores in a column, e.g., PTC($r$,1) $\forall r\in[R]$.
The partial sum from cores within one row (tile) will be accumulated via photocurrent summation.
Therefore, we can only remap the workloads in the granularity of tiles.
Formally, we denote the indices of tiles as $\mathcal{V}=[v_1, v_2,\cdots, v_{R}]$ and the indices of weight chunks as $\mathcal{U}=[u_1,u_2,\cdots,u_{R}]$, where $W_{u_r}\in\mathbb{R}^{k\times Ck}$.
Tile remapping is formulated as a linear assignment problem (LAP),
\begin{equation}
\small
    \label{eq:Remapping}
    \min_{f}\sum_{u\in\mathcal{U}} \epsilon(u, f(u)),~v=f(u)\in\mathcal{V},~f:\mathcal{U}\rightarrow \mathcal{V}.
\end{equation}
Each entry in the cost matrix $\epsilon\in\mathbb{R}^{R\times R}$ represents the edge weight in the complete bipartite graph, shown in Fig.~\ref{fig:TileRemapping}.
The edge weight $\epsilon_{ij}$ is an indicator of errors when mapping $W_i$ to tile $j$.
Similar to the salience scores in the calibration, we also use the first-order Taylor expansion to calculate sensitivity-aware error terms $\epsilon_{ij}=|\nabla_{W_i}\calL^T\cdot(\E[\widetilde{W}_i^j]-W_i^*)|$.
This LAP can be optimally solved in polynomial time.
The cycle cost of remapping one $Rk\times Ck$ weight matrix block is,
\begin{equation}
    \label{eq:Matching}
    \begin{aligned}
        \#\texttt{Cycle}_{remap}&=\#\texttt{Cycle}_{\epsilon}+\#\texttt{Cycle}_{LAP}\approx Rmk+R^3,
    \end{aligned}
\end{equation}
where the probing times $m=1$.
We periodically solve this optimal remapping $f$ to save cost and apply it to all following inferences.

\subsection{Adaptive Remediation Controller}
To dynamically determine when to trigger our remediation flow, we introduce an adaptive controller that periodically monitors cheap but informative statistics, i.e., temperature per PTC, and determine whether to trigger remediation based on a threshold, e.g., when average chip temperature drift from last remediation is above 0.01K, i.e., ($\frac{1}{RC}\sum T_{r,c}-T_{prev}>0.01$).
If not, we will perform a more expensive probing, the normalized MAE (NMAE) $\|\widetilde{W}-W^*\|_1/\|W^*\|_1$, with a slight cycle overhead.
If the NMAE is above 5\%, we will trigger remediation.
To avoid overly frequent remediation that keeps interrupting the online inference stream, we set a cooling time $\tau$ for our remediation procedure to control the maximum acceptable overhead.
This can be reconfigured by users based on the preferences between accuracy and inference throughput.
For example, with 10K total inferences, each remediation is as expensive as 10 inferences, and a cooling time of 200 inferences will lead to a max overhead of $\frac{10K\times 10}{200\times 10K}=5\%$.

\section{Evaluation Results}
\label{sec:ExperimentalResults}
\subsection{Simulation Setup}
\label{sec:ExpSetup}
\noindent\textbf{Dataset and Models.}~ 
We evaluate our method on a three-layer CNN (C64K3-C64K3-C64K3-Pool5-FC10) on Fashion-MNIST\cite{NN_FashionMNIST2017}, VGG-8\cite{NN_2015_Simonyan} on CIFAR-10\cite{NN_cifar2009}, and ResNet-18\cite{NN_CVPR2016_He} on CIFAR-100\cite{NN_cifar2009} for image classification.

\noindent\textbf{Training Settings.}~
We pre-train all models for 100 epochs with an Adam optimizer with a 2E-3 learning rate, a cosine decay scheduler, 1E-4 weight decay, and data augmentation (random crop and flip). BatchNorm layers are all frozen after pretraining.

\noindent\textbf{Architecture Settings.}~
As a challenging case study, the hardware platform in this work is assumed to be a multi-tile, multi-core photonic tensor accelerator based on a thermally sensitive MRR weight bank shown in Fig.~\ref{fig:ArchSetting}. 
Note that our method is not specific to MRR weight banks but can \emph{generalize to all universal optical matrix-vector multiplication units}.
We assume that the photonic accelerator has 4 tiles, and each tile has 4 cores. 
Each core is an 8$\times$8 add-drop MRR weight bank that can perform 8$\times$8 matrix-vector multiplication per core per cycle.  
Detailed architecture description is in Section \ref{sec:ArchSetting}.
The MRR device modeling is based on Section~\ref{sec:PhotonicsBasics}. The device/circuit variation modeling is based on Section~\ref{sec:ThermalVariationModeling}.

\noindent\textbf{Benchmark Settings}.~
To cover different thermal variation scenarios, we create several synthetic noise configurations as benchmarks in Table~\ref{tab:Benchmark}.
\begin{table}[h]
    \centering
    \caption{\small Benchmark settings for different noise scenarios}
    \resizebox{0.95\columnwidth}{!}{
    \begin{tabular}{c|c}
        \toprule
        Scenario & Description \\
        \midrule
        PV.1 & Low Noise \& Distribution: Edge-to-Corner \\ 
        \midrule
        PV.2 & High Noise \& Distribution: Edge-to-Corner\\
        \midrule
        TD.1 & Temp Drift: Linear Increase \& Uniform\\
        \midrule
        TD.2 & Temp Drift: Cosine Fluctuation \& Uniform \\
        \midrule
        TD.3 & Temp Drift: Linear Increase \& Corner Hotspot\\
        \midrule
        TD.4 & Temp Drift: Cosine Fluctuation \& Corner Hotspot\\\midrule
        CT   & Thermal Crosstalk among all MRRs within each core.\\
        \bottomrule
    \end{tabular}
    }
    \label{tab:Benchmark}
\end{table}

\textbf{(1) Phase Variation (PV)}.~To simulate real-world chip noise scenarios, two cases are created: "Low Noise" and "High Noise," corresponding to low and high standard deviation (std.) in phase shift, modeled as $\Delta\phi\in\mathcal{N}(0, \sigma^2)$. 
As explained in Section~\ref{sec:ThermalVariationModeling}, the noise std. $\sigma_{ij}^{t+1'}$ for each MRR device is time-variant and sampled at each time step from a noise level map $\mathcal{N}\in(\mu_s(t), \sigma_s^2(t))$. 
The noise level distribution gradually changes its high-noise region from the chip's left edge to the top-left corner ("Edge-to-Corner") to capture dynamic noise profile drifting. Specific noise level functions are chosen for the Low and High Noise cases:
\begin{itemize}
    \item Low Noise: $\mu_s(t)=0.0025t, \sigma_s(t)=0.004t+0.002$
    \item High Noise: $\mu_s(t)=0.01t, \sigma_s(t)=0.005t+0.005$
\end{itemize}
The damping factor $\beta_{\sigma}$ is set to 0.9.
The phase error intensities adopt the typical values in the literature~\cite{NP_DATE2020_Gu,NP_OE2019_Fang}.

\textbf{(2) Temperature Drift (TD)}: Two cases are designed to represent different types of temperature change with $t_{max}=20,000$:
\begin{itemize}
    \item Linear: $T(t)=300K + t/t_{max}$
    \item Cosine: $T(t)=300.25K - 0.25K \cos(10t/t_{max})$
\end{itemize}
The temperature drifts adopt typical values in the literature~\cite{NP_TVLSI2013_Ye,NP_Nanophotonics2014_Padmaraju}.
We also consider two representative spatial distributions of temperature.
\begin{itemize}
    \item Uniform: All PTCs have the same temperature drift following the Linear/Cosine scheduling.
    \item Corner Hotspot: the upper-left corner of the chip experiences a high temperature drift, exponentially decreasing with distance, i.e., $T(t)\gets e^{-\sqrt{r^2+c^2}}(T(t)-T(0))+T(0)$, which represents a local hotspot scenario during the execution of the accelerator.
\end{itemize}

\textbf{(3) Thermal Crosstalk (CT)}:, we use a default MRR spacing $l_v=200 \mu m$, $l_h=60 \mu m$, crosstalk coefficient $k_1=0.1$, and model crosstalk only among MRRs within the same PTC~\cite{NP_JLT2015_Jayatilleka,NP_JLT2019_Milanizadeh,NP_IPC2023_Cem}.

\noindent\textbf{Evaluation Metrics}.~
We mainly evaluate different methods on the above benchmarks in terms of inference accuracy and cycle overhead consumed by remediation.
All matrix multiplication operations in the forward propagation of NN inference is mapped to our photonic accelerators as the basic inference cycle count.
If on-chip remediation method is applied, any additional computation will be converted to equivalent \emph{cycle overhead}.
For \doctor, cycle overhead of calibration and remapping for each $Rk\times Ck$ matrix block is counted as Eq.~\eqref{eq:CalibrationCost} and Eq.~\eqref{eq:Matching}, respectively. 
The overhead is summed over all matrix blocks in the ONN.
Cycle overhead is reported as an efficiency metric for each remediation method.

\subsection{Ablation Study}
\subsubsection{Calibration: Sparsity and Salience Scores}
\label{sec:AblationSparsity}

\begin{figure}
    \centering
    \includegraphics[width=\columnwidth]{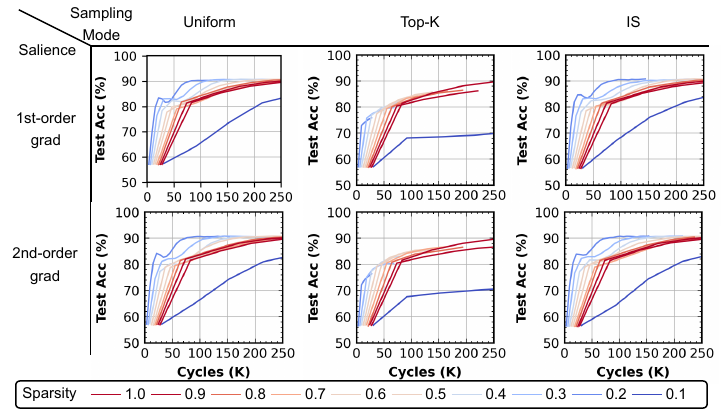}
    \caption{\small Evaluate salience, sampling methods, and sparsity levels when calibrating VGG-8 CIFAR10.
    \emph{IS} means importance sampling.}
    \label{fig:Sparsity}
\end{figure}

\begin{figure}
    \centering
    \includegraphics[width=0.85\columnwidth]{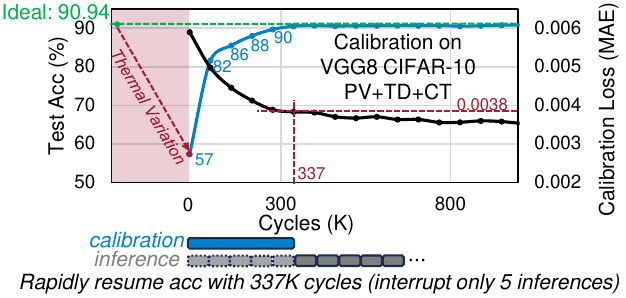}
    \caption{\small Visualization of \emph{in-situ} calibration to quickly resume accuracy at low overhead (only interrupt 5 inferences) under time-variant thermal noises on VGG8 CIFAR10.
    We adopt MAE calibration loss, learning rate of 2e-3, sparsity of 1, and averaging times $m$=1.}
    \label{fig:Mapping_Validate}
\end{figure}

To determine the best hyperparameters in sparse calibration, we first evaluate how many matrix probings $m$ are needed for Eq.~\eqref{eq:AverageProbe}. 
In Table~\ref{tab:AverageProbes}, we found that $m=1$ can efficiently estimate $\E[\widetilde{W}]$ and calibrate the circuits with low cost and high accuracy.

\begin{table}[htp]
\centering
    \caption{\small Comparison of VGG8 accuracy and calibration cycles on CIFAR-10 with different weight probing times $m$ in our proposed sparse calibration. The stop MAE threshold is 0.0038.}
    \resizebox{\linewidth}{!}{
    \begin{tabular}{c|cccccccc}
    \toprule
   $m$ & 1                               & 2                                        & 3                               & 4                               & 5                               & 10                              & 15                              & 20                              \\ \midrule
    cycles        & \textbf{3.07M}   & 1.53M   & 2.12M   & 2.82M   & 3.53M   & 6.75M   & 10.12M  & 13.50M  \\
    MAE loss      & \textbf{0.00388} & 0.00379 & 0.00382 & 0.00367 & 0.00361 & 0.00366 & 0.00360 & 0.00360 \\
    acc           & \textbf{90.67\%} & 90.05\% & 89.17\% & 89.36\% & 89.33\% & 88.56\% & 88.25\% & 88.32\% \\ \bottomrule
    \end{tabular}
    }
    \label{tab:AverageProbes}
\end{table}

Figure~\ref{fig:Sparsity} compares different salience scores, sampling methods, and sparsity.
First-order Taylor expansion with important sampling achieves the best balance between cycle counts and accuracy.
Top-K is not satisfying as it is fixed to blocks with large ideal gradients.
With sparsity $\beta=0.2$, the accuracy can be quickly resumed above 90\% with negligible cycle overhead (equivalent to one single-image inference).

Figure~\ref{fig:Mapping_Validate} visualizes the effectiveness of our salience-aware sparsity calibration.
With thermal variation, the accuracy drops from 90.94\% to 57\%.
Calibration can quickly resume accuracy above 90\% with only 10 iterations.
The overhead is very cheap, equivalent to interrupting merely 5 single-batch inferences.
With 0.2 sparsity, the overhead can be further reduced to only one inference.

\subsubsection{Remapping: Error Estimation and Interval}

\begin{figure}
    \centering
    \includegraphics[width=0.7\columnwidth]{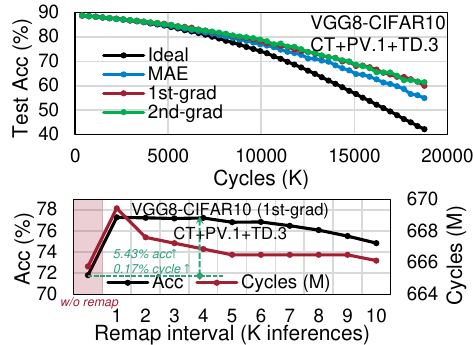}
    \caption{Using the first-order error as the sensitivity-aware $\Sigma$ and performing remapping every 4000 inferences lead to 5.4\% higher accuracy with only 0.17\% total cycle overhead.}
    \label{fig:RemappingAblation}
\end{figure}

Figure~\ref{fig:RemappingAblation} evaluates different methods for error $\epsilon_{ij}$ estimation, including mean absolute errors (MAE), first-order and second-order Taylor expansions, as indicated in Eq.~\eqref{eq:LossApproximation}.
As noise distribution evolves, the ideal pre-trained model with direct mapping suffers from a large accuracy drop.
In contrast, our variation-aware remapping can significantly reduce the numerical errors and thus boost inference accuracy by 5-10\%.
The first-order Taylor expansion of the loss function shows clear advantages over the naive matrix MAE scores and is cheaper than second-order expansion with the same accuracy benefit.
The solved remapping can be reused for following inferences until the next remapping is triggered.
We scan over different remapping intervals from once per 1K inferences up to once per 10K inferences.
We found that 2K-4K inference intervals are enough to guarantee maximum accuracy benefits (+5.4\%) and negligible cycle overhead (0.17\%).

\subsection{Main Results: Compare with Prior Work}
\label{sec:MainResults}
\begin{table}[]
\caption{Compare the inference accuracy and cycle overhead (\emph{Ovhd}) among 4 methods on 8 different variation settings and 3 datasets/models.
Our remediation method only induces a small overhead of 0.14\%-5\% of the original inference cycles, which is 2-3 orders-of-magnitude less costly than on-chip training (BS3).}
\label{tab:MainResults}
\resizebox{\columnwidth}{!}{
\begin{tabular}{c|cccccccc}
\hline
                                                               & \multicolumn{8}{c}{FMNIST-CNN3 (pre-trained acc: 92.78\%, inference cycles: 5.81E8)}                                                        \\ \cline{2-9} 
\multirow{-2}{*}{Noise configs}                                & \multicolumn{2}{c}{BS1} & \multicolumn{2}{c}{BS2~\cite{NP_DATE2020_Gu}} & \multicolumn{2}{c}{BS3~\cite{NP_NeurIPS2021_Gu, NP_Science2023_Pai}} & \multicolumn{2}{c}{\cellcolor[HTML]{EFEFEF}\doctor}               \\
                                                               & acc        & ovhd       & acc    & ovhd      & acc        & ovhd       & \cellcolor[HTML]{EFEFEF}acc   & \cellcolor[HTML]{EFEFEF}ovhd   \\ \hline
CT+PV.1+TD.1                                                      & 50.41      & 0          & 45.56       & 0         & 36.72      & 3.5E9      & \cellcolor[HTML]{EFEFEF}92.42 & \cellcolor[HTML]{EFEFEF}8.4E5  \\
CT+PV.1+TD.2                                                      & 42.85      & 0          & 37.08       & 0         & 77.74      & 3.5E9      & \cellcolor[HTML]{EFEFEF}92.59 & \cellcolor[HTML]{EFEFEF}7.9E5  \\
CT+PV.1+TD.3                                                      & 78.44      & 0          & 81.49       & 0         & 80.50      & 3.5E9      & \cellcolor[HTML]{EFEFEF}92.20 & \cellcolor[HTML]{EFEFEF}7.6E5  \\
CT+PV.1+TD.4                                                      & 74.11      & 0          & 75.09       & 0         & 90.37      & 3.5E9      & \cellcolor[HTML]{EFEFEF}92.19 & \cellcolor[HTML]{EFEFEF}7.5E5  \\
CT+PV.2+TD.1                                                      & 49.43      & 0          & 45.39       & 0         & 27.49      & 3.5E9      & \cellcolor[HTML]{EFEFEF}91.42 & \cellcolor[HTML]{EFEFEF}8.4E5  \\
CT+PV.2+TD.2                                                      & 41.81      & 0          & 37.00       & 0         & 74.64      & 3.5E9      & \cellcolor[HTML]{EFEFEF}91.07 & \cellcolor[HTML]{EFEFEF}8.2E5  \\
CT+PV.2+TD.3                                                      & 77.94      & 0          & 80.55       & 0         & 78.14      & 3.5E9      & \cellcolor[HTML]{EFEFEF}91.12 & \cellcolor[HTML]{EFEFEF}8.4E5  \\
CT+PV.2+TD.4                                                      & 73.34      & 0          & 74.20       & 0         & 88.54      & 3.5E9      & \cellcolor[HTML]{EFEFEF}91.11 & \cellcolor[HTML]{EFEFEF}8.4E5  \\ \hline
\begin{tabular}[c]{@{}c@{}}Avg. Acc/\\ Ovhd Ratio\end{tabular} & 61.04      & 0.00\%     & 59.55       & 0.00\%    & 69.27      & 600\%      & \cellcolor[HTML]{EFEFEF}91.77 & \cellcolor[HTML]{EFEFEF}0.14\% \\ \hline
\end{tabular}
}

\vspace{10pt}
\resizebox{\columnwidth}{!}{
\begin{tabular}{c|cccccccc}
\hline
                                                               & \multicolumn{8}{c}{CIFAR10 VGG8 (pre-trained acc: 90.94\%, inference cycles: 6.66E8)}                                                        \\ \cline{2-9} 
\multirow{-2}{*}{Noise configs}                                & \multicolumn{2}{c}{BS1} & \multicolumn{2}{c}{BS2~\cite{NP_DATE2020_Gu}} & \multicolumn{2}{c}{BS3~\cite{NP_NeurIPS2021_Gu, NP_Science2023_Pai}} & \multicolumn{2}{c}{\cellcolor[HTML]{EFEFEF}\doctor}               \\
                                                               & acc        & ovhd       & acc        & ovhd       & acc        & ovhd       & \cellcolor[HTML]{EFEFEF}acc   & \cellcolor[HTML]{EFEFEF}ovhd   \\ \hline
CT+PV.1+TD.1                                                      & 32.38      & 0          & 29.93      & 0          & 19.23      & 3.3E9      & \cellcolor[HTML]{EFEFEF}90.53 & \cellcolor[HTML]{EFEFEF}2.6E7  \\
CT+PV.1+TD.2                                                      & 28.54      & 0          & 26.86      & 0          & 53.71      & 3.3E9      & \cellcolor[HTML]{EFEFEF}90.24 & \cellcolor[HTML]{EFEFEF}2.4E7  \\
CT+PV.1+TD.3                                                      & 71.87      & 0          & 67.12      & 0          & 73.11      & 3.3E9      & \cellcolor[HTML]{EFEFEF}90.25 & \cellcolor[HTML]{EFEFEF}2.4E7  \\
CT+PV.1+TD.4                                                      & 65.82      & 0          & 60.89      & 0          & 86.47      & 3.3E9      & \cellcolor[HTML]{EFEFEF}90.27 & \cellcolor[HTML]{EFEFEF}2.6E7  \\
CT+PV.2+TD.1                                                      & 32.52      & 0          & 30.00      & 0          & 19.17      & 3.3E9      & \cellcolor[HTML]{EFEFEF}90.25 & \cellcolor[HTML]{EFEFEF}2.6E7  \\
CT+PV.2+TD.2                                                      & 28.98      & 0          & 27.02      & 0          & 51.96      & 3.3E9      & \cellcolor[HTML]{EFEFEF}89.62 & \cellcolor[HTML]{EFEFEF}2.4E7  \\
CT+PV.2+TD.3                                                      & 72.03      & 0          & 67.26      & 0          & 71.68      & 3.3E9      & \cellcolor[HTML]{EFEFEF}89.85 & \cellcolor[HTML]{EFEFEF}2.4E7  \\
CT+PV.2+TD.4                                                      & 65.80      & 0          & 60.87      & 0          & 85.83      & 3.3E9      & \cellcolor[HTML]{EFEFEF}89.75 & \cellcolor[HTML]{EFEFEF}2.3E7  \\ \hline
\begin{tabular}[c]{@{}c@{}}Avg. Acc/\\ Ovhd Ratio\end{tabular} & 49.74      & 0.00\%     & 46.24      & 0.00\%     & 57.65      & 500\%      & \cellcolor[HTML]{EFEFEF}90.10 & \cellcolor[HTML]{EFEFEF}3.65\% \\ \hline
\end{tabular}
}

\vspace{10pt}
\resizebox{\columnwidth}{!}{
\begin{tabular}{c|cccccccc}
\hline
                                                               & \multicolumn{8}{c}{CIFAR100 ResNet18 (pre-trained acc: 73.57\%, inference cycles: 5.43E9)}                                                   \\ \cline{2-9} 
\multirow{-2}{*}{Noise configs}                                & \multicolumn{2}{c}{BS1} & \multicolumn{2}{c}{BS2~\cite{NP_DATE2020_Gu}} & \multicolumn{2}{c}{BS3~\cite{NP_NeurIPS2021_Gu, NP_Science2023_Pai}} & \multicolumn{2}{c}{\cellcolor[HTML]{EFEFEF}\doctor}               \\
                                                               & acc        & ovhd       & acc        & ovhd       & acc        & ovhd       & \cellcolor[HTML]{EFEFEF}acc   & \cellcolor[HTML]{EFEFEF}ovhd   \\ \hline
CT+PV.1+TD.1                                                      & 5.23       & 0          & 5.32       & 0          & 6.20       & 2.7E10     & \cellcolor[HTML]{EFEFEF}72.15 & \cellcolor[HTML]{EFEFEF}2.9E8  \\
CT+PV.1+TD.2                                                      & 7.25       & 0          & 7.66       & 0          & 24.03      & 2.7E10     & \cellcolor[HTML]{EFEFEF}71.33 & \cellcolor[HTML]{EFEFEF}2.8E8  \\
CT+PV.1+TD.3                                                      & 21.69      & 0          & 23.06      & 0          & 39.00      & 2.7E10     & \cellcolor[HTML]{EFEFEF}70.20 & \cellcolor[HTML]{EFEFEF}2.7E8  \\
CT+PV.1+TD.4                                                      & 18.16      & 0          & 18.82      & 0          & 46.56      & 2.7E10     & \cellcolor[HTML]{EFEFEF}70.01 & \cellcolor[HTML]{EFEFEF}2.6E8  \\
CT+PV.2+TD.1                                                      & 5.14       & 0          & 5.36       & 0          & 4.96       & 2.7E10     & \cellcolor[HTML]{EFEFEF}72.28 & \cellcolor[HTML]{EFEFEF}2.9E8  \\
CT+PV.2+TD.2                                                      & 7.33       & 0          & 7.65       & 0          & 24.90      & 2.7E10     & \cellcolor[HTML]{EFEFEF}71.68 & \cellcolor[HTML]{EFEFEF}2.8E8  \\
CT+PV.2+TD.3                                                      & 21.97      & 0          & 23.61      & 0          & 1.03       & 2.7E10     & \cellcolor[HTML]{EFEFEF}70.47 & \cellcolor[HTML]{EFEFEF}2.7E8  \\
CT+PV.2+TD.4                                                      & 18.43      & 0          & 19.21      & 0          & 46.27      & 2.7E10     & \cellcolor[HTML]{EFEFEF}70.40 & \cellcolor[HTML]{EFEFEF}2.6E8  \\ \hline
\begin{tabular}[c]{@{}c@{}}Avg. Acc/\\ Ovhd Ratio\end{tabular} & 13.15      & 0.00\%     & 13.84      & 0.00\%     & 24.12      & 500\%      & \cellcolor[HTML]{EFEFEF}71.07 & \cellcolor[HTML]{EFEFEF}5.08\% \\ \hline
\end{tabular}
}
\end{table}

In Table~\ref{tab:MainResults}, we compare \doctor with three baselines: (1) BS1: deploy pre-trained models; (2) BS2~\cite{NP_DATE2020_Gu}: noise-aware training with (2\%) weight error injected during pretraining; and (3) BS3~\cite{NP_NeurIPS2021_Gu, NP_Science2023_Pai}: on-chip training for 1 epoch on a calibration dataset (10\% of training set).
The parameters for our \doctor framework are: (1) \emph{Calibration}: probing times $m=1$, salience score $s=|\nabla_{W}\calL|$, calibration sparsity $\beta=0.2$, max calibration iteration $T_{calib}=20$ (50 for ResNet18); 
(2) \emph{Remapping}: first-order Taylor expansion as $\epsilon$;
(3) \emph{Controller}: remediation cooling time $\tau$: 200 inferences (50 for ResNet18); trigger remapping $\rightarrow$ calibration. 

\begin{figure}
    \centering
    \includegraphics[width=0.9\columnwidth]{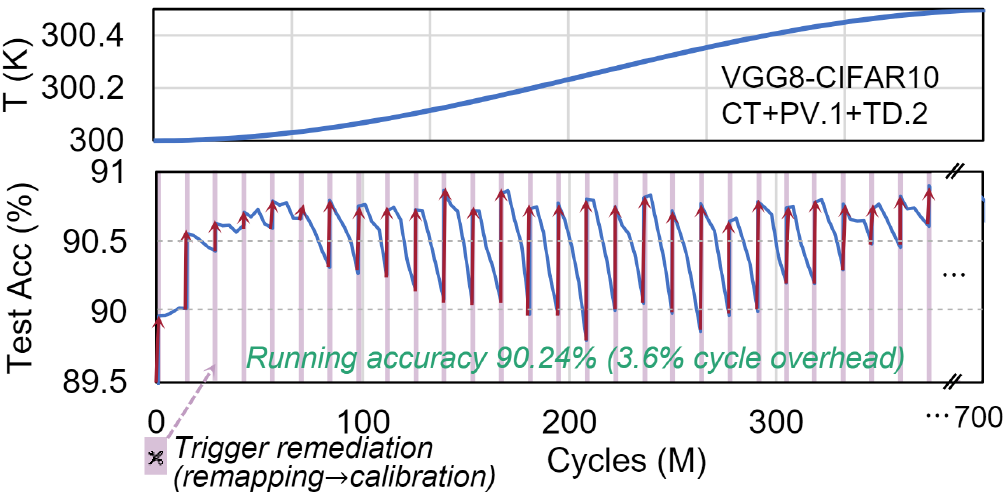}
    \caption{\small Visualize \doctor for dynamic accuracy recovery.}
    \label{fig:Visualize}
\end{figure}

Across all benchmarks, BS1 suffers from severe accuracy degradation (30\%-60\%).
Noise-aware training~\cite{NP_DATE2020_Gu}, though it enhances the smoothness of the solution space of the model, shows limited effect on accuracy improvement, as the noise distribution used in pre-training is significantly different from the physical variations.
Note that our \doctor is orthogonal to noise-aware training, allowing simultaneous application for a solution space that is both locally smooth and adaptive to the drifting noise distribution.
On-chip training~\cite{NP_NeurIPS2021_Gu, NP_Science2023_Pai} can boost the accuracy on small networks, e.g., 8\% on CNN-FMNIST and VGG8-CIFAR10. 
However, it performs poorly on deep ResNet18 and is not stable on certain benchmarks, e.g., CT+PV.2+TD.1.
The fundamental reason is that the gradient estimation error exponentially accumulates with backpropagation, which can lead to poor training performance and even divergence.
Also, though it is only trained for 1 epoch on a 10\% training set, it consumes 5-6$\times$ more runtime (cycles) in training, which is not practical to be deployed on throughput-restricted edge platforms.
In contrast, our proposed \doctor method can stably maintain high accuracy even with rapidly drifting noise distributions and temperature with \textbf{less than 1-2.5\% accuracy drop} at the cost of merely \textbf{0.1\%-5.1\% cycle overhead}.
On average, \doctor is +34\% more accurate and 2-3 orders-of-magnitude more efficient than on-chip training.
We visualize our \doctor flow in Fig.~\ref{fig:Visualize} to show the temperature drift is detected and the accuracy is rapidly resumed with only 3.6\% cycle overhead.

\begin{figure}[]
    \centering
    \includegraphics[width=0.75\columnwidth]{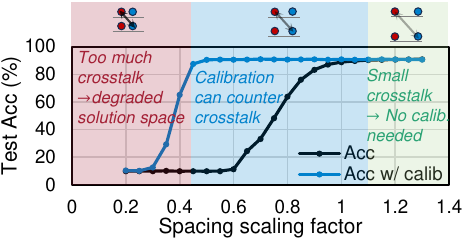}
    \caption{\small Impact of device spacings on crosstalk-induced accuracy drop and calibration effectiveness.
    We adopt 200 calibration iterations for sufficient convergence.}
    \label{fig:CrosstalkSpacing}
\end{figure}

\subsection{Discussion}

\noindent\textbf{Device Spacing and Crosstalk}.~
We evaluate how MRR spacing impacts the crosstalk and thus the maximum accuracy in Fig.~\ref{fig:CrosstalkSpacing}.
By default, we assume 200 $\mu m$ vertical spacing $l_v$ and 60 $\mu m$ horizontal spacing $l_h$ in the MRR array and scaling both directions.
Below 0.4, the crosstalk severely impacts the representable weight space. 
Thus, even after calibration, the accuracy cannot be recovered.
Within 0.4 to 1.1 scaling, the accuracy drop can be fully countered by our calibration, while above 1.1, there is almost no drop from crosstalk.

\noindent\textbf{Trade off Efficiency and Accuracy}.~
The cooling time $\tau$ in our adaptive remediation controller trades off cycle overhead and accuracy.
In Table~\ref{tab:Cooltime}, we sweep the cooling interval from 200 to 1000 inferences.
If accuracy is prioritized, we can adopt $\tau$=200 with only 3.7\% cycle overhead.
If inference throughput is prioritized, e.g., only <1\% overhead is accepted, we can set $\tau=800$ with a 2.7\% accuracy drop.
\begin{table}[htp]
\centering
\caption{\small Cycle overhead and test accuracy on VGG8-CIFAR10 with different remediation cooling time $\tau$.}
\resizebox{\columnwidth}{!}{
\label{tab:Cooltime}
\begin{tabular}{c|cccccc}
\toprule
$\tau$                                                        & 200                                                       & 300                                                      & 400                                                      & 500                                                      & 800                                                      & 1000                                                     \\ \midrule
\begin{tabular}[c]{@{}c@{}}\#cycles\\ (overhead)\end{tabular} & \begin{tabular}[c]{@{}c@{}}6.90E8\\ (+3.7\%)\end{tabular} & \begin{tabular}[c]{@{}c@{}}6.78E8\\ (+1.9\%)\end{tabular} & \begin{tabular}[c]{@{}c@{}}6.78E8\\ (+1.9\%)\end{tabular} & \begin{tabular}[c]{@{}c@{}}6.74E8\\ (+1.3\%)\end{tabular} & \begin{tabular}[c]{@{}c@{}}6.72E8\\ (+1.0\%)\end{tabular} & \begin{tabular}[c]{@{}c@{}}6.71E8\\ (+0.8\%)\end{tabular} \\
acc                                                           & 90.24                                                     & 89.97                                                    & 89.97                                                    & 89.15                                                    & 88.28                                                    & 86.55                                                    \\ \bottomrule
\end{tabular}
}
\end{table}

\vspace{-5pt}
\section{Conclusion}
\label{sec:Conclusion}
In this work, we present \emph{the first} on-chip remediation approach that dynamically monitors photonic accelerator temperature drift and ensures continued reliability with minimal overhead through training-free, data-free calibration, and architectural remapping. 
Our method outperforms SoTA on-chip training by +34\% higher accuracy and 2-3 orders-of-magnitude lower cost.
Our lightweight, effective \emph{in-situ} remediation method enables self-corrected photonic neural accelerators with unprecedented reliability in real-world, dynamic deployment scenarios.

\bibliographystyle{IEEEtran}


\begin{IEEEbiographynophoto}{Haotian Lu}
is currently a research intern in ScopeX group, School of Electrical, Computer and Energy Engineering at Arizona State University, advised by Prof. Jiaqi Gu. His research interests mainly include hardware-algorithm co-design, electronic design automation, efficient hardware accelerators and hardware security.    
\end{IEEEbiographynophoto}
\vspace{-0.2in}

\begin{IEEEbiographynophoto}{Sanmitra Banerjee}
is a Senior Design-for-X (DFX) Methodology Engineer at NVIDIA Corporation, Santa Clara, CA, and an Adjunct Faculty at Arizona State University. He received the B.Tech. degree from Indian Institute of Technology, Kharagpur, in 2018, and the M.S. and Ph.D. degrees from Duke University, Durham, NC, in 2021 and 2022, respectively. His research interests include machine learning based DFX techniques, and fault modeling and optimization of emerging AI accelerators under process variations and manufacturing defects.
\end{IEEEbiographynophoto}
\vspace{-0.2in}

\begin{IEEEbiographynophoto}{Jiaqi Gu} (S'19 - M'23)
received the B.E.~degree in Microelectronic Science and Engineering from Fudan University, Shanghai, China in 2018, and the Ph.D. degree in the Department of Electrical and Computer Engineering, The University of Texas at Austin, Austin, TX, USA in 2023. 
He is currently an Assistant Profoessor in School of Electrical, Computer and Energy Engineering at Arizona State University, Tempe, AZ, USA.
His current research interests include emerging hardware design for efficient computing (photonics, post-CMOS electronics, quantum), hardware-algorithm co-design, AI/ML algorithms, and electronic-photonic design automation.
He has received the Best Paper Award at IEEE TCAD 2021, the Best Paper Award at ASP-DAC 2020, the Best Paper Finalist at DAC 2020, the Best Poster Award at NSF Workshop on Machine Learning Hardware (2020), the ACM/SIGDA Student Research Competition First Place (2020), and the ACM Student Research Competition Grand Finals First Place (2021).
\end{IEEEbiographynophoto}
\vspace{-0.2in}

\vfill
\end{document}